\documentclass[oldversion]{aa}			

\usepackage{graphicx}
\usepackage{txfonts}

\newcommand{\ergs}{erg~cm$^{-2}$~s$^{-1}$}
\newcommand{\gsim}{{\lower.5ex\hbox{$\; \buildrel > \over \sim \;$}}}
\begin{document}

\title{Roma-BZCAT: A multifrequency catalogue of Blazars}

\author{E. Massaro\inst{1}
\and P. Giommi\inst{2}
\and C. Leto\inst{2} 
\and P. Marchegiani\inst{1} 
\and A. Maselli\inst{1}
\and M. Perri\inst{2}
\and S. Piranomonte\inst{3} 
\and S. Sclavi\inst{1}
}
	
\institute{
Dipartimento di Fisica, Sapienza Universit\`a di Roma, Piazzale A. Moro 2, I-00185, 
Roma, Italy 
\and
ASI Science Data Center, ASDC c/o ESRIN, via G. Galilei, I-00044 Frascati, Italy 
\and
INAF, Osservatorio Astronomico di Roma, Monte Porzio Catone, Italy
 }

\offprints{~~~~~~~~\\
E.~Massaro: ~enrico.massaro@uniroma1.it}

\date{Received:.; accepted:.}

\titlerunning{Roma-BZCAT: a multifrequency catalogue of blazars}
\authorrunning{E. Massaro et al.}

\abstract
{We present a new catalogue of blazars based on multi-frequency surveys
and on an extensive review of the literature. Blazars are classified
as BL Lacertae objects, as flat spectrum radio quasars or as blazars of
uncertain/transitional type. Each object is identified by a root name, 
coded as BZB, BZQ and BZU for these three subclasses respectively, 
and by its coordinates.
This catalogue is being built as a tool useful for the identification of
the extragalactic sources that will be detected by present and future 
experiments for X and gamma-ray astronomy, like Swift, AGILE, Fermi-GLAST and 
Simbol-X.
An electronic version is available from the ASI Science Data Center web site
at http://www.asdc.asi.it/bzcat.
}


\maketitle


\section{Introduction}
Blazars are a relatively rare class of active galactic nuclei (AGN) characterised 
by an electromagnetic emission over the entire energy spectrum, from the radio band to 
the most energetic $\gamma$ rays.
Angel and Stockman (1980) reported that the word BLAZAR, a combination of 
BL Lac object and quasar, was proposed by Edward A. Spiegel in a talk given at 
the banquet of the famous first BL Lac Conference in Pittsburgh (1978). 
At that epoch these sources were compact flat-spectrum radio sources with  
a violent flaring activity, featureless optical spectra and occasionally showing
a high linear polarisation, as first recognised by Strittmatter et al. (1972).
Their emission in the X and $\gamma$-ray bands was practically unknown.
This name was lucky because it was soon adopted and became official with 
the review paper by Angel and Stockman (1980). 
These authors gave also one of the first list of BL Lac objects containing 
57 sources.

The discovery that some radio-loud quasars show characteristics similar
to BL Lac objects, with the exception of broad spectral lines,
was important for a complete definition of the blazar properties.
Moore and Stockman (1981) gave an important contribution when performed
a polarisation survey in which discovered 17 high polarisation quasar
(HPQ) and discussed their relation with BL Lacs. 
Before their work only four radio quasar were known to show high variable 
optical linear polarisation.
Some years later Impey and Tapia (1988) extended the polarisation
surveys and discovered 31 new blazars.

The unusual properties of blazars were explained at the Pittsburgh 
Conference by Blandford and Rees (1978) in terms of non-thermal emission 
from small regions moving down a jet with a velocity close to the 
speed of light and observed at a small angle with respect to the line 
of sight.
The Doppler boosting that results from these very special circumstances 
strongly amplifies the non-thermal emission that dominates with 
respect to all other components of thermal origin, like those from the 
accretion disk, the dust heated by the AGN flux, or from stars 
and interstellar matter of the host galaxy.

Many surveys have been carried out to identify blazars and their number
is continuously increasing.
Nevertheless, only a relative small number of blazars have been studied 
intensively.
There are still many unsolved problems about the origin of the blazar 
phenomenon, the physical processes occurring in the nuclear region, and 
about the cosmological evolution of these sources.
Furthermore, because of their very broad spectral emission, blazars are 
expected to play an important role in generating the extragalactic 
background and of its fluctuations in several energy bands.
 
Blazars constitute the most numerous class of extragalactic $\gamma$-ray 
sources detected by the EGRET experiment onboard Compton-GRO (Hartman et al.
1999), and by atmospheric Cerenkov telescopes at TeV energies. 
Moreover, a recent analysis of the arrival directions of cosmic rays with
energy above $\sim$6$\times$10$^{19}$ eV collected at the Auger observatory,
showed a significant correlation with the coordinates of AGNs within $\sim$
75 Mpc (Pierre Auger Collaboration 2007).
The knowledge of the blazar population is, therefore, very relevant in 
high-energy astrophysics.

For these reasons a complete list of known blazars, based on an accurate examination 
of literature data, can be very useful for carrying out some of the scientific 
goals of present and future space observatories like Fermi-GLAST, Planck, and Simbol-X.
In this paper we describe the ``Multifrequency Catalogue of BLAZARS", also named 
{\it Roma-BZCAT}, that will be published in four volumes (the first two are
already available; Massaro et al. 2005, Massaro et al. 2008) each of them 
containing a sky region covering 6 hours of right ascension.
This catalogue is a master list of sources useful for the search of counterparts
in different spectral bands and for the selection of samples for statistical studies 
of blazar properties and evolution.
An electronic version is also available on-line from the ASDC web page
(http://www.asdc.asi.it/bzcat),
where it is updated as new objects are reported. 

\section{Blazar surveys}

Starting from the late 1980's, several sizable blazar samples were produced 
as a result of a number of systematic searches in large radio or X-ray surveys.
These highly organised efforts were initially carried out following a standard
approach involving the optical identification of all sources above the survey's 
flux limit.
However, when the need to assemble deeper and larger samples pushed the 
need for optical telescope time to hardly manageable amounts, 
multi-frequency (usually radio, optical, and X-ray), pre-selection techniques 
were developed in order to significantly reduce the number of candidates requiring 
optical spectroscopy.

After the first list of 32 BL Lac objects given by Stein, O'Dell, and Strittmatter 
(1976), Angel and Stockman (1980) increased the number to 57. 
A catalogue of BL Lac objects was prepared by Burbidge and Hewitt (1987), 
containing 87 sources, but it did not include all the sources listed by Angel and 
Stockman. 
At that epoch the total number of blazar was 103.
Since 1984 Veron-Cetty and Veron published a general catalogue of AGN that, starting 
from the 2nd edition (Veron-Cetty and Veron 1985), included a table of BL Lac objects, 
whose number is continuously increasing from the original 73 to 1,122 of the 12th edition 
(Veron-Cetty and Veron 2006). 
Although very useful, the VCV catalogue is not a blazar catalogue, because many sources 
are listed in the BL Lac table and others among the quasars, and for this reason, the 
blazar nature of the sources is not directly apparent.

\begin{table*}[h]
\caption{Major Blazar surveys}
\begin{center}
\begin{tabular}{lllcc}
\hline
                &   &          &      &  \\
{Survey name}   & Blazar type  &Survey type  &  Number  & Reference \\
                &              &             &  of objects &  \\
\hline
                &   &          &      &  \\
  2Jansky       & BL Lacs+FSRQs  &  radio flux limited  & 60 & 1, 2, 3 \\
                &                &  $F(2.7~{\rm GHz}) >$ 2 Jy, $\alpha_r \le 0.4$ & &  \\ 
  1Jansky       & BL Lacs  &  radio flux limited & 34 & 4 \\
                &          & $F(5~{\rm GHz}) >$ 1 Jy, $\alpha_r \le 0.5$ & & \\ 
 1Jy NVSS-RASS  & BL Lacs+FSRQs& radio, opt, X-ray selection &  157 & 5 \\
                &              & radio fl. limited, $F(1.4~{\rm GHz}) >$ 1 Jy  &                    &  \\
 1Jy WMAP       & BL Lacs+FSRQs& microwave flux limited &180 & 6, 7 \\
                &              & $F(61~{\rm GHz}) \gsim$ 1 Jy     & &  \\  
 Parkes 0.25 Jy & BL Lacs+FSRQs& radio flux limited&371    & 8,9 \\
                &              & $F(2.7~{\rm GHz}) >$ 0.25 Jy           &                    &  \\
 EMSS           & BL Lacs  &   X-ray flux limited & 41 & 10, 11 \\
                &          & $F_{X} \gsim 2\times 10^{-13}$ \ergs  &                    &  \\  
 IPC slew survey& BL Lacs  & X-ray flux limited   &  51& 12   \\
                &          & $F_{X} \gsim 2\times 10^{-12}$ \ergs &                    &  \\ 
 DXRBS          & BL Lacs+FSRQs&radio+opt+X-ray selection &  227 & 13, 14  \\
                &              & radio fl. limited &                    &  \\
                &                & $F(5~{\rm GHz}) > $ 50 mJy & &  \\ 
  REX            & BL Lacs      & radio+opt+X-ray selection & $\sim$100  & 15, 16  \\
 RGB            & BL Lacs      & radio+opt+X-ray selection   & 127  & 17 \\
 Sedentary      & extreme BL Lacs &radio+opt+X-ray selection&  150   & 18, 19 \\
                &              &radio fl. limited  &  &  \\
                &              & $F(1.4~{\rm GHz}) >$ 3.5 mJy  &           &  \\
 CLASS          & BL Lacs+FSRQs& radio+opt &   147 & 20, 21 \\
                &              & $F(5~{\rm GHz}) >$ 30 mJy, $\alpha_r \le 0.5$ &  &  \\
 Gamma-ray Blazars & BL Lacs+FSRQs & radio+X-ray selection &  241 & 22 \\
 2dF BL Lac Survey & BL Lacs  & Optical selection &  56 candidates & 23, 24  \\
 SDSS BL Lac Survey & BL Lacs  & Optical selection &  347 & 25 \\
  ROXA            & BL Lacs+FSRQs+AGN & radio+opt+X-ray selection   & 816  & 26 \\
  CGRaBS          & BL Lacs+FSRQs+AGN & radio flux limited & 1625  & 27 \\
                &   &          &      &  \\
\hline

\end{tabular}

\end{center}
\scriptsize

(1) Wall and Peacock (1985), 
(2) di Serego-Alighieri et al. (1994),  
(3) Urry, Padovani (1995), 
(4) Stickel et al. (1991), 
(5) Giommi et al. (2002), 
(6) Bennett C.L. et al. (2003) 
(7) Giommi, Colafrancesco, (2004),
(8) jackson et al. (2002),  
(9) Wall et al. (2005),  
(10) Stocke et al. (1991),
(11) Rector et al. (2000),
(12) Perlman et al. (1996),
(13) Perlman et al. (1998), 
(14) Landt et al. (2001), 
(15) Caccianiga et al. (1999),
(16) Caccianiga et al. (2002a),  
(17) Laurent-Muehleisen et al. (1998),
(18) Giommi, Menna, Padovani (1999), 
(19) Giommi et al. (2005),
(20) March\~a et al. (2001), 
(21) Caccianiga et al. (2002b), 
(22) Sowards-Emmerd et al. (2005), 
(23) Londish et al. (2002), 
(24) Londish et al. (2007),
(25) Collinge et al. (2005),
(26) Turriziani et al. (2007),
(27) Healey et al. (2008).
\end{table*}

A sample oriented catalogue was compiled by Padovani \& Giommi (1995) 
including all BL Lacertae objects that were known at that time. After 
ten years that list has become largely incomplete and needs updating.   

In recent years new surveys have become available and the number of blazar 
candidates has grown rapidly. 
An important multifrequency sample is the radio-optical-X-ray built at
ASDC ({\it ROXA}, Turriziani et al. 2007), which was obtained by means of
a cross-correlation between large radio ({\it NVSS}, {\it ATCAPMN}
Taker 2000) and X-ray surveys ({\it RASS}), together with the {\it SDSS-DR4} 
and {\it 2dF} data to spectroscopically identify the candidates. 
The {\it ROXA} sample consists of 816 objects and includes 173 newly 
discovered blazars. 
All the ROXA objects were examined to verify that
they satisfy the {\it Roma-BZCAT} acceptance criteria, and only
a small percentage of them was not accepted. 
The relatively high threshold of the X-ray flux of the RASS
survey mainly selects objects with high $F_X/F_{radio}$ flux
ratio, hence HBL objects.

Another sample of BL Lac candidates, including 386 sources, was
obtained by Collinge et al. (2005) using the {\it SDSS} spectroscopic 
database in a field of 2860 deg$^2$.
These sources were divided into the two groups of 240 ``probable" and 146 
``possible" BL Lac candidates, the former selected by colours, unlike those of
DC white dwarf stars, and, some evidence of extragalactic nature, such as
a redshift estimate and/or a radio/X-ray counterpart.
The majority of these BL Lac candidates are, as for the {\it ROXA} sample, 
of the HBL type. 
In this sample there are several sources without a radio counterpart,
and this possibility raises the issue of the existence of radio-quiet
(or optically selected) BL Lacs.
We are already faced with this problem when considering the sample of 
``optically selected" BL Lac candidates by Londish et al. (2002),
as a byproduct of the much wider {\it 2QZ} sample of QSOs (Croom et al. 
2001).
Nesci et al. (2005) searched for optical variability of a subsample 
of the {\it 2QZ-BL} sources and found that only a few of them have significant 
brightness changes, including all the radio-loud ones. 
In a recent paper, Londish et al. (2007) reported a further analysis
of an expanded and revised sample of potential optically identified
BL Lacs and concluded that there can be no significant population of
radio-quiet BL Lac objects.

Although in principle we cannot exclude the existence of high-$z$ HBL
objects with a sub mJy flux density in the GHz band, we preferred to 
maintain our criterion to include in the {\it Roma-BZCAT} only those sources 
with a confirmed or high confidence radio counterpart;
however, in Table 2 of the 12th edition of their AGN catalogue 
(V\'eron-Cetty and V\'eron 2006) reported all the BL Lac candidates given
in the above samples, therefore including also radio-quiet objects, 
but clearly indicating that they are not confirmed.

A new important survey aimed at indentifying new blazars is
the Candidate Gamma Ray Blazar Survey ({\it CGRaBS}) (Healey et al. 2008).
This sample is essentially based on the {\it CRATES} (Combined Radio All-Sky
Targeted Eight GHz Sources) catalogue (Healey et al. 2007), which 
extended the {\it CLASS} (Cosmic Lens All-Sky Survey, Myers et al. 2003)
to obtain observations at 8.4 GHz of flat spectrum sources with 
$|b|~> 10^{\circ}$ and brighter than 65 mJy at 4.8 GHz, with the
exclusion of the region with declination $\delta > 75^{\circ}$ where
the flux limit is 250 mJy.
The {\it CRATES} catalogue contains 11,131 sources, many of them without
an optical counterpart.

The {\it CGRaBS} sample was selected to identify sources with the 
greatest similarity to the EGRET blazars, and an optical follow-up 
provided the counterpart identification and spectroscopy for the majority 
of them.
For the sake of completeness, we decided to add all {\it CGRaBS} sources 
classified as FSRQs to the {\it Roma-BZCAT}, although their spectra are
not available in the literature and/or on-line.
These sources are marked by the note CG.
Sources of unknown type and without redshift were not included.

Table 1 gives a list of all the major surveys of both types that led to the discovery 
of most of the blazars known today and included in this catalogue.

\begin{figure*}
\resizebox{\hsize}{!}
{\includegraphics[width=5cm,angle=-90]{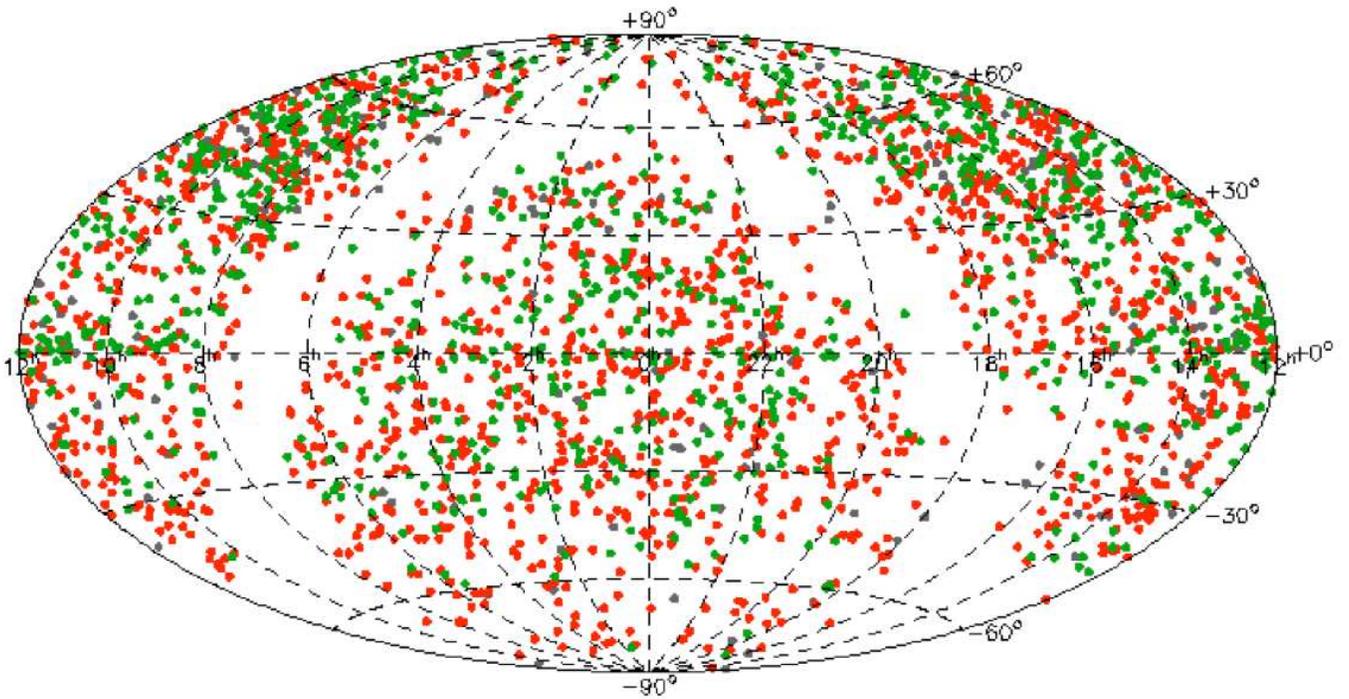}}
\caption{Aitoff projection in equatorial coordinates showing the sky 
distribution of blazars listed in the on-line version of {\it Roma-BZCAT}. 
Red points indicate BZB sources, green points BZQ objects, and blue points BZU
sources; the empty belt corresponds to the Milky Way and the most crowded region 
in the northern emisphere to the SDSS field.
}
\label{RBZfig1}
\end{figure*}

\section{Criteria for source selection}
As stated in the introduction, blazars come in two main flavours: BL Lac objects 
(BL) and Flat Spectrum Radio Quasars (FSRQ).
The former type is characterised by featureless optical spectra, while 
those of FSRQs display the prominent emission lines that 
are typical of QSOs.
This classification, although very simple, is not unambiguous.
For instance, well-established BL Lac objects like OJ 287 and  BL Lac itself 
on some occasions clearly exhibited emission lines, indicating common nature 
of the two types of blazars.
Another problem concerns variability: sometimes blazars showed large brightness 
changes on time scales of a day or of a few hours, whereas it is possible that
they remain in a quasi-quiescent state for years. 
Such a behaviour can hide blazar properties and can lead to misidentification.
In very general terms we can define a blazar as an extragalactic source 
characterised by 
$i$) a spectral energy distribution (SED) showing strong non thermal emission 
over the entire electromagnetic spectrum, from radio waves to $\gamma$-rays, 
and 
$ii$) the presence of some evidence 
for relativistic beaming.
It is clear that both requirements can be secured by only making use of 
rich observational data sets, involving either polarimetric or 
variability studies and VLBI imaging to detect compact cores and 
superluminal motions.
These data sets are available only for a limited number of, usually 
bright, sources, and therefore the majority of known blazars have been 
recognized only on the basis of a few spectral properties.
Moreover, the data available are not simultaneous in most cases and 
are often derived from catalogues, sometimes making it very difficult to 
establish whether an AGN is a genuine blazar or not.

For these reasons we adopted a rather wide set of acceptance criteria as 
described below: 

1. detection in the radio band, down to mJy flux densities at
1.4 GHz ({\it NVSS}, Condon et al. 1998 or {\it FIRST}, White et al. 1997) 
or 0.84 GHz ({\it SUMSS}, Mauch et al. 2003);

2. optical identification and knowledge of the optical spectrum to 
establish the type of blazar;

3. isotropic X-ray luminosity close to or higher than 10$^{43}$ erg s$^{-1}$;

4. for FSRQs a spectral index ($F(\nu) \propto \nu^{-\alpha}$), 
measured between 1.4 GHz (or 0.843 GHz) and 5 GHz, $\alpha_r < 0.5$, but not
for BLs, although most of them have flat spectra;

5. compact radio morphology, or, when extended, with one dominant core
and a one-sided jet.  

Recent observational campaigns with the {\it Swift-XRT} telescope on samples of 
BL Lacs and FSRQs without measured X-ray fluxes confirmed that they emit X rays 
with an intensity in agreement with that of similar blazars (Giommi et al. 2007).

Some authors have proposed lists of optically selected BL Lacs on the basis of featureless
spectra without the occurrence of a radio and X-ray counterpart 
(e.g. 2dF BL Lac survey).
These sources are reported in some widely used catalogues, although with a doubtful
indication. 
To reduce the possibility of including non blazar sources in the catalogue, we
adopted a conservative criterion and inserted only those sources in the table of 
candidate BL Lacs having well-established radio emission.

Another problem concerns the nature of a class of radio sources whose nature
is still a matter of debate:
the so called High-Frequency Peakers (HFP) (Dallacasa et al. 2000)
and GHz-Peaked Sources (GPS) (O'Dea et al. 1991), whose optical counterparts
are a mixture of galaxies and quasars, together with a few BL Lacs.
Broad-band monitoring campaigns have shown that some of these sources are 
indeed variable and that their spectra can change from the characteristic 
convex shape to a flat frequency distribution (Torniainen et al. 2005).  
Such behaviour suggests a link between GPS-HFP objects and blazars.
Moreover, the spectral coverage of many radio sources is poor, both in frequency
and time, and several sources with non simultaneous spectra could have been 
classified as FSRQ, while they are GPS-HFP and vice-versa.
In other cases, some SEDs indicate the occurrence of a convex radio component
superposed onto a power-law spectrum.
We decided therefore to include this type of radio sources in the catalogue,
having blazar characteristics and adding a specific note if they are GPS or HFP, 
using the recent master list by Labiano et al. (2007) and other 
literature as reference.    

\section{General organisation of the {\it Roma-BZCAT}}
The {\it Roma-BZCAT} is structured in two parts.
Part 1, at present the only one available on-line, contains the lists of blazars, 
which are classified in three main groups, according to their spectral properties.
Each blazar is identified by a three-letter code, where the first two are {\bf BZ}
for blazar and the third one specifies the type, followed by the truncated 
equatorial coordinates (J2000).

The codes are:
\begin{itemize}
\item
{\bf BZB}: BL Lac objects, used for AGNs with a featureless optical spectrum or
only with absorption lines of galaxian origin and weak and narrow emission lines;

\item
{\bf BZQ}: flat-spectrum radio quasars, with an optical spectrum showing broad emission
lines and dominant blazar characteristics;

\item
{\bf BZU}: blazars of uncertain type, adopted for sources with  
peculiar characteristics but also showing blazar activity: for instance, occasional
presence/absence of broad spectral lines or features, transition objects between a
radio galaxy and a BL Lac, galaxies hosting a low luminosity blazar nucleus, etc.;

\end{itemize}

The BZB objects are listed in the first two tables of the catalogue (but they are not
distinguished in the on-line version).
Table I contains all firmly identified BL Lacs,
whereas Table II includes sources considered to be {\it candidate BL Lac objects}, in 
particular those for which we could not find, either in literature or on the web,
an optical spectrum or a good description.
The BZQ sources listed in Table III are FSRQs and other quasars showing evident blazar 
characteristics like a strong and fast optical variability (OVV sources) and a
high linear polarisation (HPQ sources).
Finally, we listed in Table IV the BZU sources with a short note explaining the reason
for this classification. 
 
The following data are reporte for each source:
J2000 coordinates, $R$ magnitude (from USNO-B1), radio flux density
at 1.4 GHz (from NVSS or other catalogues, if not known the flux at another frequency 
is given), the X-ray flux in the 0.1--2.4 keV energy interval, mostly from ROSAT
archive, and the redshift. 
Some useful synthetic notes are also given for a limited number of blazars:
e.g. member of a galaxy cluster, detection of emission lines, etc.

The on-line version reports the same data and offers the additional  
possibility of using the ASDC "Data Explorer" tool.
This tool is being developed at the ASDC to help search for more  
information from several on-line databases. 
In particular, for each source, it provides a direct link to NED and SIMBAD, 
which can be used to find other names, references, and other useful information.

\begin{figure}
{\includegraphics[width=6.7cm,angle=-90]{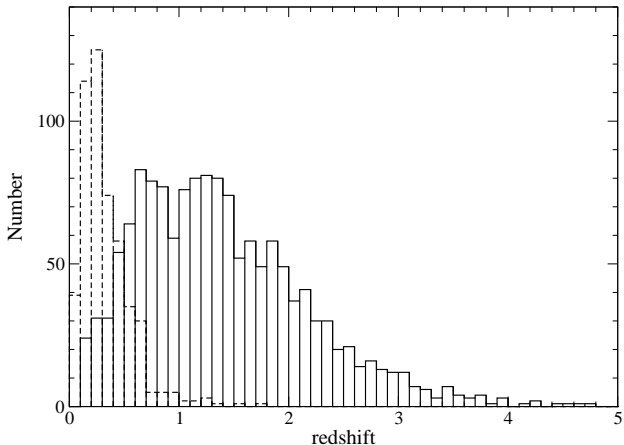}}
\caption{The redshift distributions of BL Lac objects, including
candidate sources (dashed histogram) and of FSRQs (solid line histogram)
reported in the {\it Roma-BZCAT}.
}
\label{RBZfig2}
\end{figure}

\section{Blazar population and spectral energy distributions}
The total number of blazars presently listed in the {\it Roma-BZCAT} is 2728
and their distribution in the sky is shown in Fig.~1. 
The most populated class is that of FSRQs, which are 57.1\% of sources.
BL Lac objects (including candidates) are 33.9\%, and the remaining 9.0\% are sources 
of uncertain classification.
Only about the 42\% of FSRQs are detected in the X-ray band.  
A large observational work is then necessary to complete the knowledge
of the emission from FSRQs in this band.

A relevant and well known difference between the populations of FSRQs and BL objects
is the redshift distributions as shown by the histograms given in Fig. 2.  
BL Lacs are much closer than FSRQs and their distribution peaks at $z \simeq 0.3$,
whereas quasars show a broad maximum between 0.6 and 1.5. 
There are only very few BL objects at redshift higher than 0.8, but these 
estimates are  generally very uncertain, and the extension of the distribution 
in this range cannot be considered fully reliable. 
One must also to take into account that z is unknown for about 40\% of BZB sources
and that there is a bias towards near and low power sources whose galaxian lines
are easily detectable.

A section of the {\it Roma-BZCAT} is devoted to presenting detailed data and
other useful information for a selection of interesting blazars to represent 
the various blazar types and with special attention to those detected in the 
$\gamma$- and hard X-ray ranges.   
In particular, for each object we build the SED using a
large collection of non-simultaneous multi-frequency data.
Although SEDs of extragalactic sources are indeed given in the NED database, we made
an effort to enrich fluxes derived from catalogues by adding more data taken from
papers and observational databases.
In the case of monitoring campaigns containing a lot of data we considered only the 
lowest and highest flux levels.
An example, the SED of the well known BZB J0721+7120 (S5 0716+714, 3EG J0712+7120),
is shown in Fig. 3.

\begin{figure}
{\includegraphics[height=8.4cm,angle=-90]{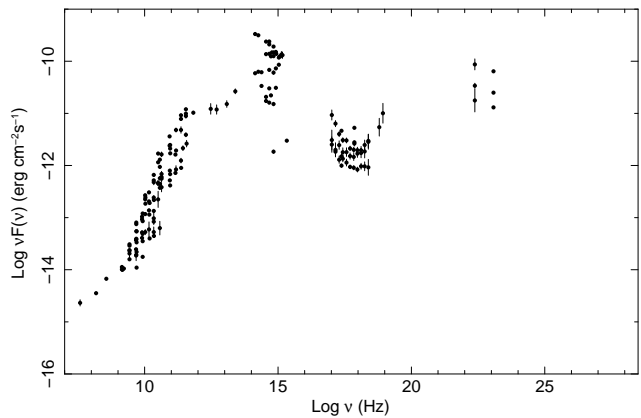}}
\caption{The SED of the BL Lac object BZB J0721+7210 (S5 0716+714) 
derived from the {\it Roma-BZCAT} database.
}
\label{RBZfig3}
\end{figure}

\section{Conclusion}
The main scientific goals that led us to the compilation of the {\it Roma-BZCAT} were

$i$) to have the most complete list of published blazars, useful for identifying 
the counterparts of high-energy sources (in this respect the Roma-Bzcat is already one 
of the reference catalogues of the Fermi-GLAST mission);

$ii$) to have a list of extragalactic objects useful for studying the populations of 
high-energy extragalactic sources;

$iii$) to have a population from which it will be possible to extract samples 
satisfying statistical criteria for investigating blazar properties and evolution;

$iv$) to have a large database of SEDs for different types of blazars useful for
studying radiation mechanisms and relativistic beaming effects.

An open problem is that of the completeness of the catalogue. 
There is, in fact,
a deficiency of blazars in the southern sky because of the fewer surveys
than the northern emisphere.
Many new discovered blazars and candidates, for instance, have been identified
using spectroscopic observations available in the {\it SDSS}.
A similar research project for exploring the region around the southern celestial
pole, possibly coupled with a high sensitivity radio survey, will certainly lead to
the discovery of hundreds of new blazars. 
Moreover, we expect that the deeper search for faint optical counterparts of flat 
spectrum radio sources will surely increase the number of blazars at high redshift 
and provide new useful information for undertstanding their cosmological evolution.
To take such new possible developments  into account, we will continue in the next
years to update the on-line version of the {\it Roma-BZCAT} and to revise the data 
and, eventually, the classification of sources. 

\begin{acknowledgements}
This work has made extensive use of the following on-line facilities:   
the National Extragalactic Database (NED),
The Astrophysics Data System (ADS), SIMBAD, NRAO calibrator database, ASDC.
We acknowledge the financial support by Agenzia Spaziale Italiana
(ASI-INAF n. I/010/06/0) for astrophysical science activity for the Fermi-GLAST mission.
We thank Roberto Primavera for his help in the preparation of the on- 
line version of the catalogue.
We are also grateful to the referee, M.P. V\'eron-Cetty, for helpful comments.
\end{acknowledgements}

{}

\end{document}